\documentclass[aps,prb,showpacs,twocolumn,superscriptaddress]{revtex4-1}

\usepackage{graphicx}
\usepackage{braket}
\usepackage{amsmath}
\usepackage{amssymb}
\usepackage{color}
\usepackage{soul}
\usepackage[squaren,Gray]{SIunits}

\usepackage{notes2bib}

\begin{document}


\title{Localization landscape theory of disorder in semiconductors II: Urbach tails of disordered quantum well layers}


\author{Marco Piccardo}
\email[]{marco.piccardo@polytechnique.edu}
\affiliation{Laboratoire de Physique de la Mati\`ere Condens\'ee, Ecole polytechnique, CNRS, Universit\'e Paris Saclay, 91128 Palaiseau Cedex, France}
\affiliation{Materials Department, University of California, Santa Barbara, California 93106, USA}

\author{Chi-Kang Li}
\affiliation{Graduate Institute of Photonics and Optoelectronics and Department of Electrical Engineering, National Taiwan University, Taipei 10617, Taiwan}

\author{Yuh-Renn Wu}
\affiliation{Graduate Institute of Photonics and Optoelectronics and Department of Electrical Engineering, National Taiwan University, Taipei 10617, Taiwan}

\author{James S. Speck}
\affiliation{Materials Department, University of California, Santa Barbara, California 93106, USA}

\author{Bastien Bonef}
\affiliation{Materials Department, University of California, Santa Barbara, California 93106, USA}

\author{Robert M. Farrell}
\affiliation{Materials Department, University of California, Santa Barbara, California 93106, USA}

\author{Marcel Filoche}
\affiliation{Laboratoire de Physique de la Mati\`ere Condens\'ee, Ecole polytechnique, CNRS, Universit\'e Paris Saclay, 91128 Palaiseau Cedex, France}

\author{Lucio Martinelli}
\affiliation{Laboratoire de Physique de la Mati\`ere Condens\'ee, Ecole polytechnique, CNRS, Universit\'e Paris Saclay, 91128 Palaiseau Cedex, France}

\author{Jacques Peretti}
\affiliation{Laboratoire de Physique de la Mati\`ere Condens\'ee, Ecole polytechnique, CNRS, Universit\'e Paris Saclay, 91128 Palaiseau Cedex, France}

\author{Claude Weisbuch}
\affiliation{Laboratoire de Physique de la Mati\`ere Condens\'ee, Ecole polytechnique, CNRS, Universit\'e Paris Saclay, 91128 Palaiseau Cedex, France}
\affiliation{Materials Department, University of California, Santa Barbara, California 93106, USA}


\date{\today}

\begin{abstract}
Urbach tails in semiconductors are often associated to effects of compositional disorder. The Urbach tail observed in InGaN alloy quantum wells of solar cells and LEDs by biased photocurrent spectroscopy is shown to be characteristic of the ternary alloy disorder. The broadening of the absorption edge observed for quantum wells emitting from violet to green (indium content ranging from 0 to 28\%) corresponds to a typical Urbach energy of 20~meV. A 3D absorption model is developed based on a recent theory of disorder-induced localization which provides the effective potential seen by the localized carriers without having to resort to the solution of the Schr\"odinger equation in a disordered potential. This model incorporating compositional disorder accounts well for the experimental broadening of the Urbach tail of the absorption edge. For energies below the Urbach tail of the InGaN quantum wells, type-II well-to-barrier transitions are observed and modeled. This contribution to the below bandgap absorption is particularly efficient in near-UV emitting quantum wells. When reverse biasing the device, the well-to-barrier below bandgap absorption exhibits a red shift, while the Urbach tail corresponding to the absorption within the quantum wells is blue shifted, due to the partial compensation of the internal piezoelectric fields by the external bias. The good agreement between the measured Urbach tail and its modeling by the new localization theory demonstrates the applicability of the latter to compositional disorder effects in nitride semiconductors. 
\end{abstract}


\maketitle

\section{Introduction}

The absorption edge in semiconductors is usually characterized by a long energy tail with an absorption coefficient~$\alpha$ exponentially varying with photon energy and temperature along the Urbach tail rule~\cite{Urbach1953}
\begin{equation}
\alpha\left(h\nu\right) = \alpha_0 \cdot e^{(h\nu - E_1)/E_U(T)}~,
\end{equation}
where $\alpha_0$ is a constant, $E_1$ determines the onset of the tail, usually the bandgap energy $E_g$ in semiconductors, and $E_U$ is the Urbach energy. Note that Urbach tails were proposed in 1953 essentially as a phenomenological description of thermally activated below-gap absorption (see e.g. Ref.~\citenum{Kurik1971} and references therein) and in Urbach's original expression $E_U = kT/\gamma$, $\gamma$ being a constant. Several attempts have since then been made trying to derive a universal theory.\cite{Dow1972, John1986} By extension, exponential absorption edges are also called Urbach tails although they might not be associated with thermally activated phenomena. A variety of phenomena, intrinsic or extrinsic, have been invoked, and sometimes demonstrated, as physical mechanisms leading to Urbach tails such as the Franz-Keldysh (FK) effect associated with the random electric fields of impurities, defect-induced smearing of band edges~\cite{Hwang1970} and phonon-assisted absorption.\cite{Kurik1971} The impact of compositional alloy disorder on thermally activated Urbach tails has been clearly identified in InAs$_x$Sb$_{1-x}$~\cite{Bansal2007} while in the case of amorphous Si Cody et al. could characterize both the effects of structural disorder and thermal disorder.\cite{Cody1981}

In the InGaN alloy system, very large Urbach energies $E_U$ of few hundreds of meV have been observed,\cite{Han2006, Bayliss1999, ODonnell1999} attributed to phase separation of the InGaN alloy. This is to be contrasted with the luminescence linewidth in LED materials which is most often quite smaller, of the order of 100~meV or less. The Urbach energy observed in pure GaN is 12~meV~\cite{Chichibu1997} or 15~meV.\cite{Jacobson2001} As will be shown in the present paper, the Urbach tail of high quality In$_x$Ga$_{1-x}$N quantum wells (QWs) has Urbach energies in the range 15-30~meV and is determined by the disorder introduced by the indium compositional fluctuations.

Understanding disorder phenomena in nitride ternary alloys, such as InGaN and AlGaN, is of utmost importance as they constitute the core of several modern optoelectronic devices due to their wide tunability of energy bandgap. It is indeed well established by atom probe tomography (APT) that in these alloys atoms are randomly distributed and thus induce random potential fluctuations.\cite{Wu2012} Although APT provides imaging of the compositional disorder with near atomic resolution, the relation between the measured atomic maps and the resulting energy fluctuations occurring near the band gap edge of the alloy, where all physical phenomena of devices occur, is an area of ongoing research -- including the work reported here.

Theoretical and numerical studies indicate that compositional disorder in nitride ternary alloys is large enough to induce carrier localization which, in turn, may strongly influence carrier transport, optical properties and efficiency of a device. In the case of nitride light-emitting diodes (LEDs), simulations of I-V~characteristics including random alloy fluctuations in the InGaN QWs and AlGaN barriers predict a voltage threshold which is almost 1~V lower with respect to the case of a homogeneous active region, in better agreement with the experimental values for actual devices, due to percolation in the disordered layers.\cite{Yang2014, Wu2015, Li2017} Two companion papers, referred to as LL1 (Ref.~\citenum{Filoche2017}) and  LL3~(Ref.~\citenum{Li2017}), describe the basis of the localization landscape (LL) theory and its application to transport and recombination in nitride heterostructure LEDs, respectively. Atomistic calculations by Schulz et al.~\cite{Schulz2015} show that low-band electron and hole states produce a strong inhomogeneous broadening of luminescence spectra of In$_{0.25}$Ga$_{0.75}$N/GaN QWs. First-principle calculations of non-radiative Auger rates in nitride semiconductors by Kioupakis et al.~\cite{Kioupakis2015} indicate that indirect Auger recombination mediated by alloy disorder may play an important role in determining the efficiency loss of nitride optoelectronic devices.

On the experimental side, besides the Urbach tail measurement mentioned above, a well-known effect often interpreted as an evidence of carrier localization induced by disorder in InGaN QW layers is the ``S-shaped'' emission shift of the peak energy of photoluminescence observed as a function of temperature.\cite{Cho1998, Schomig2004} This phenomenon indicates relaxation of the carrier ensemble at low temperatures towards deeply localized centers, signaled by a red-shift of the emission energy, but does not provide information on the disorder effects impacting device operation.

In contrast, optical absorption spectroscopy should provide a more direct access to the disorder-induced energy fluctuations near the band edge, as it is directly linked to the electronic densities of states. In the present study the role of compositional disorder in a series of In$_x$Ga$_{1-x}$N QW structures of different indium content is characterized by means of biased photocurrent spectroscopy (BPCS). Assuming that the collection efficiency of photocurrent is independent of photon energy, BPCS measurements are representative of optical absorption spectra.\cite{Helmers2013} For instance, in GaAs/AlGaAs QWs a series of distinct excitonic peaks could be observed by Collins et al.\cite{Collins1986} using BPCS, indicating that excitons, after photocreation, are ionized and transport through the structure contributing to photocurrent. To the extent that the optical interband matrix element has a weak dependence on photon energy in the narrow energy range of the Urbach tail, then below-gap absorption is determined by the joint density of states (JDOS) of the disordered system weighted by the overlap integral between the envelope functions of the localized states.\cite{Singh2007} The contribution of internal piezoelectric fields on the absorption spectra can be evaluated by comparing data collected with different external applied biases. Previous photocurrent studies were carried out in InGaN/GaN QW structures under applied bias but did not aim at characterizing disorder-related effects on below-gap absorption.\cite{Yee2010, Vierheilig2009}

In the present work we are able to characterize the effect of compositional disorder in InGaN layers with different indium content by analyzing their Urbach tails as measured by BPCS. QWs emitting from the violet ([In]=11\%) to the green ([In]=28\%) range exhibit small values of the Urbach energy, typically around 20~meV, while a much larger broadening would be expected from the fluctuations of the calculated random potential corresponding to the alloy disorder. An independent absorption model is derived on the basis of a recent mathematical theory of localization that takes into account the ``effective'' confining potential seen by localized states in disordered systems without having to resort to the Schr\"odinger equation.\cite{Filoche2012, Arnold2016} A development of the LL~theory in the framework of semiconductors is reported in LL1 (Ref.~\citenum{Filoche2017}) and its predictions account well for the experimental broadening of the below-gap absorption edge.

The description of the samples and details on the experimental setup are given in Sec.~\ref{sec:experiment}. In Sec.~\ref{sec:results} the main results of BPCS for In$_x$Ga$_{1-x}$N QW structures with indium composition ranging between 0\% and 28\% will be presented. It will be shown that three different regions may be identified in the photocurrent spectra corresponding to different absorption mechanisms. The focus will be on the experimental characterization of Urbach tails as a function of indium composition, applied bias, sample temperature and type of device incorporating InGaN QWs. In Sec.~\ref{sec:theor_model} theoretical models will be derived to study the effect of electric fields and compositional disorder on Urbach tails. First the influence of the polarization-induced QW electric field on the optical transition matrix element by means of the FK effect at below-gap photon energies will be evaluated. Then the absorption model based on the LL~theory will be introduced and its predictions will be compared to the measured Urbach tails giving an insight into the effect of compositional disorder in InGaN layers. Finally calculations of type-II well-to-barrier transitions will be presented in the Appendix to model an additional absorption mechanism observed in the below-gap responsivity of near-UV emitting QWs.

\section{Experiment}
\label{sec:experiment}

\subsection{Samples}

\begin{figure}
\includegraphics[width=0.48\textwidth]{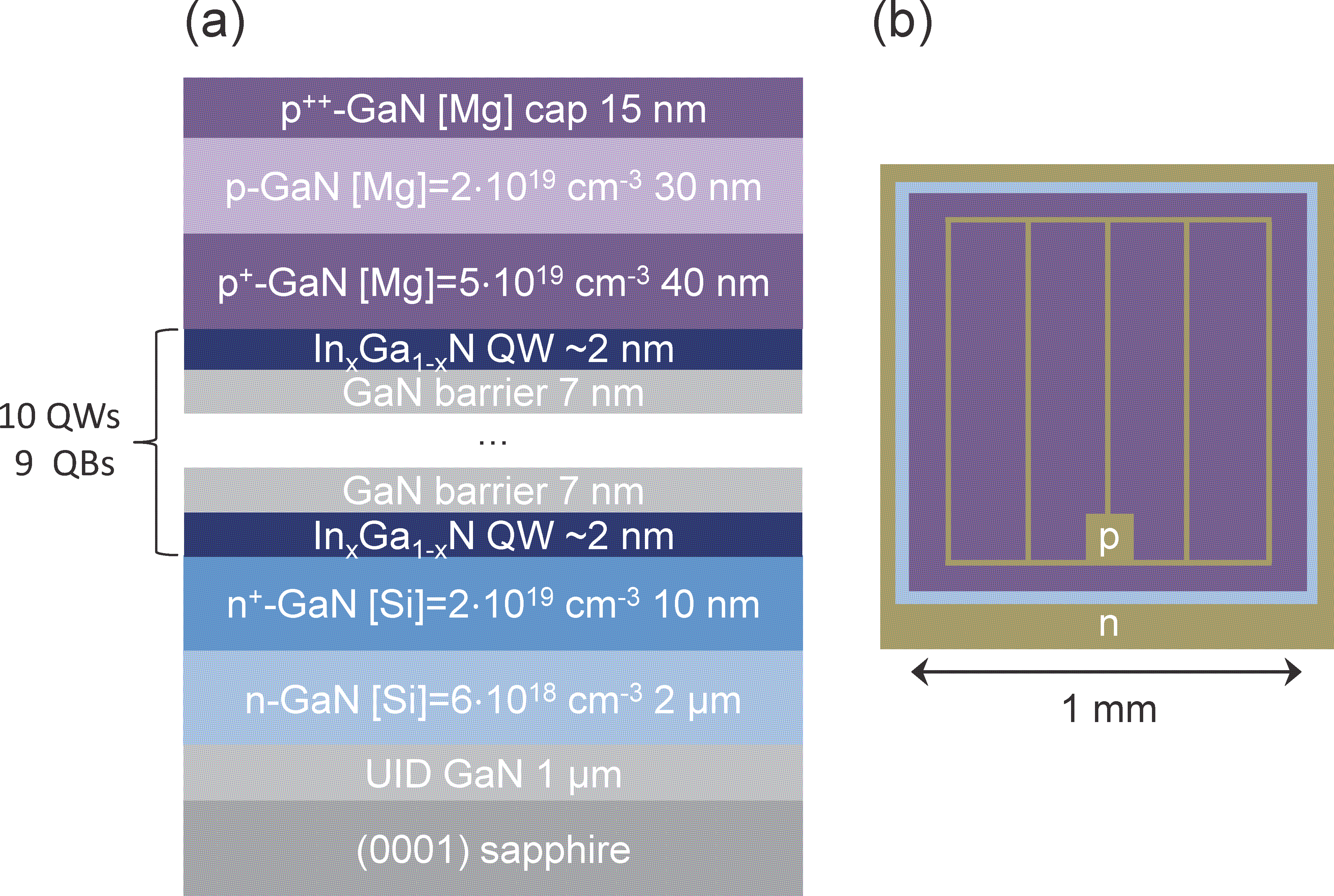}
\caption{(a) Schematic of the InGaN/GaN solar cell structures grown by MOCVD. (b) Contact geometry of the processed devices.}
\label{fig:schematic}
\end{figure}

The samples are a series of In$_x$Ga$_{1-x}$N/GaN multiple QW (MQW) solar cells grown by metal-organic chemical-vapor deposition (MOCVD) on (0001) sapphire substrates with mean indium content ranging between 6\% and 28\% measured by x-ray diffraction (XRD) with a PANalytic MRD PRO diffractometer. The structures are schematized in Fig.~\ref{fig:schematic}(a) and consist of a 1~\micro m unintentionally doped GaN template layer, a 2~\micro m Si-doped n-GaN layer ([Si]=$6 \times 10^{18}$~cm$^{-3}$), a 10~nm highly Si-doped n$^+$-GaN layer ([Si]=$2 \times 10^{19}$~cm$^{-3}$), a 10~period undoped MQW active region with $\sim$~2~nm In$_x$Ga$_{1-x}$N QWs (the QW thickness measured by XRD slightly increases with [In] in the different solar cells: 1.6~nm for 6\%, 1.7~nm for 11\%, 1.8~nm for 17\%, 1.9~nm for 22\% and 2.1~nm for 28\%) and 7~nm GaN barriers, a 40~nm highly Mg-doped p$^+$-GaN layer ([Mg]=$5 \times 10^{19}$~cm$^{-3}$), a 30~nm Mg-doped p-GaN layer ([Mg]=$2 \times 10^{19}$~cm$^{-3}$) and a 15~nm highly Mg-doped p$^+$-GaN contact layer. The n- and p- layer adjacent to the active region are $\delta$-doped to reduce the polarization fields in the QWs. APT experiments were performed to confirm the random distribution of indium atoms in the InGaN layers in the different samples. Specimens were prepared with a FEI Helios 600 dual beam FIB instrument following standard procedure. APT analyses were performed with a Cameca 3000X HR Local Electrode Atom Probe (LEAP) operated in laser-pulse mode with a green laser (532~nm). The laser pulse energy was 1~nJ and a detection rate of 0.005~atoms per pulse was set during each sample analysis. APT 3D reconstruction was carried out using commercial software IVASTM. Statistical distribution analysis were performed on sampling volumes of 30 $\times$ 30 $\times$ 2~nm$^3$ taken in the center of each InGaN layers and for each sample. The random distribution of indium in each InGaN layer was confirmed by the success of the $\chi^2$ test between experimental distributions and the binomial distribution predicted for random alloys.

A UV LED with pure GaN QW layers is also studied for comparison. This sample is grown by molecular beam epitaxy (MBE) on a semipolar ($20\bar{2}1$) GaN substrate and its structure consists of an AlGaN buffer layer, a Si-doped n-Al$_{0.20}$Ga$_{0.80}$N layer, a 5~period undoped MQW active region with 5~nm GaN QWs and 6~nm Al$_{0.12}$Ga$_{0.88}$N barriers, a Mg-doped p- Al$_{0.20}$Ga$_{0.80}$N layer and a highly Mg-doped p$^+$-GaN contact layer. The samples are processed by standard contact lithography and the final devices consist of 1~mm $\times$ 1~mm mesas, 30/300~nm Pd/Au p-contact grids on the top of each mesa with a center-to-center grid spacing of 200~\micro m, and a 30/300~nm Al/Au n-contacts around the base of each mesa [Fig.~\ref{fig:schematic}(b)].

For comparison, commercial-grade LEDs produced by Nichia Corporation are also investigated: a UV LED (NCSU275 U395), in which the glass window has been removed to avoid partial absorption in the UV range of the exciting optical beam in BPCS measurements and in which the Zener diode has been removed to apply a wider range of reverse biases to the device; a bluish-green LED (NCSE 119AT), in which the silicone dome and the Zener diode were not removed.\bibnote{Chemical removal of the silicone dome with Dynasolve solvent leaves the features of BPCS spectra substantially unchanged but considerably degrades their signal-to-noise ratio.}

\subsection{Experimental set-up}

The BPCS set-up consists of an Energetiq EQ-99 light source coupled to a TRIAX 180 monochromator with an output wavelength bandwidth of 2~nm (corresponding to about 15~meV in the range of interest of the present study). We note that for such value the convolution of an exponential decay with the Gaussian spectral line shape of the exciting beam does not affect the slope of the Urbach tail (i.e., the measurement of $E_U$). The spot size of the optical beam on the sample was $\sim$~1~mm$^2$, being comparable with the area of the mesas of the devices. The samples were biased with a Keithley 2400 current-voltage source. The optical beam was modulated with a chopper wheel at a frequency of 82~Hz and the photocurrent signal was measured with a Princeton Applied Research 5209 lock-in amplifier. Low-temperature measurements were carried out in a JANIS SHI-4 closed cycle cryostat.

\section{Results}
\label{sec:results}

\begin{figure}
\includegraphics[width=0.39\textwidth]{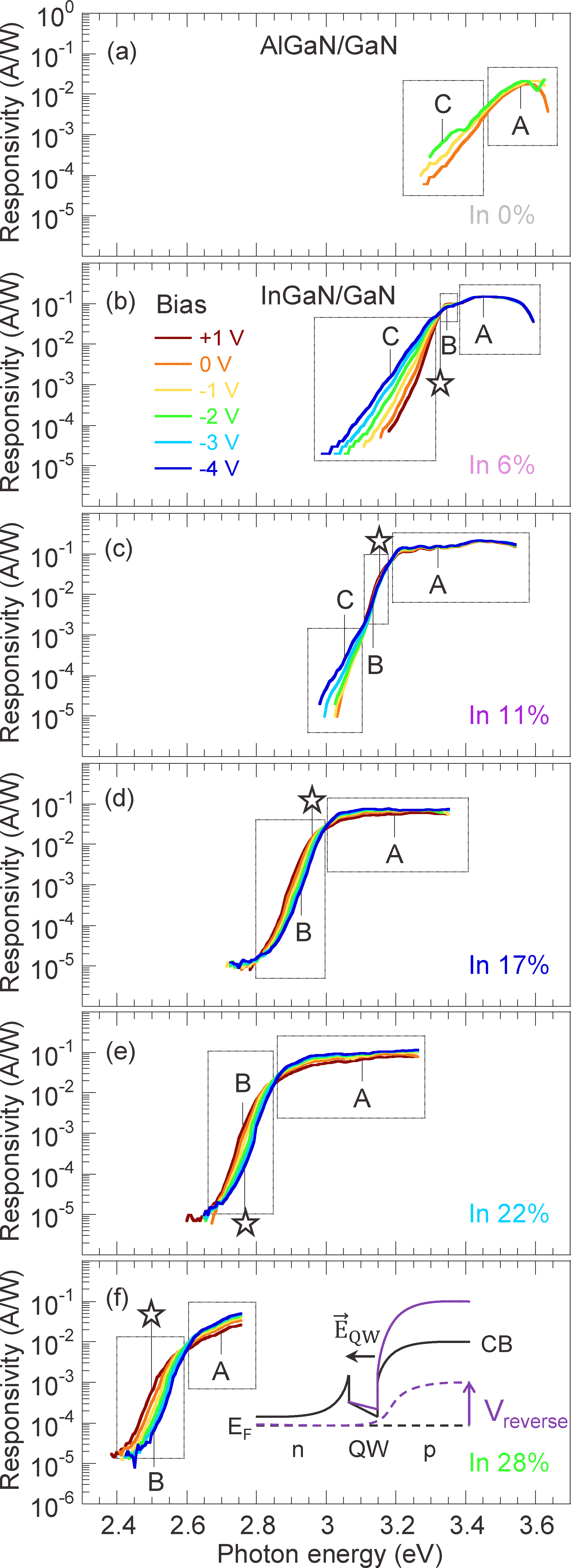}
\caption{Responsivity measurements by BPCS of (a) m-plane UV GaN/AlGaN LED, and (b)-(f) c-plane In$_x$Ga$_{1-x}$N/GaN solar cells with indium composition ranging between 6\% and 28\ The curves correspond to different applied biases to the devices (the negative sign indicates a reverse bias). Three regions having a different bias dependence and named 'A', 'B', 'C' are enclosed by boundaries. The PL peak energy of the solar cells measured at 0~V is marked as stars. The schematic in (f) shows the conduction band diagram of a c-plane InGaN/GaN QW structure in which the QW electric field is partly compensated under application of a reverse bias.}
\label{fig:responsivity1}
\end{figure}

\begin{figure}
\includegraphics[width=0.49\textwidth]{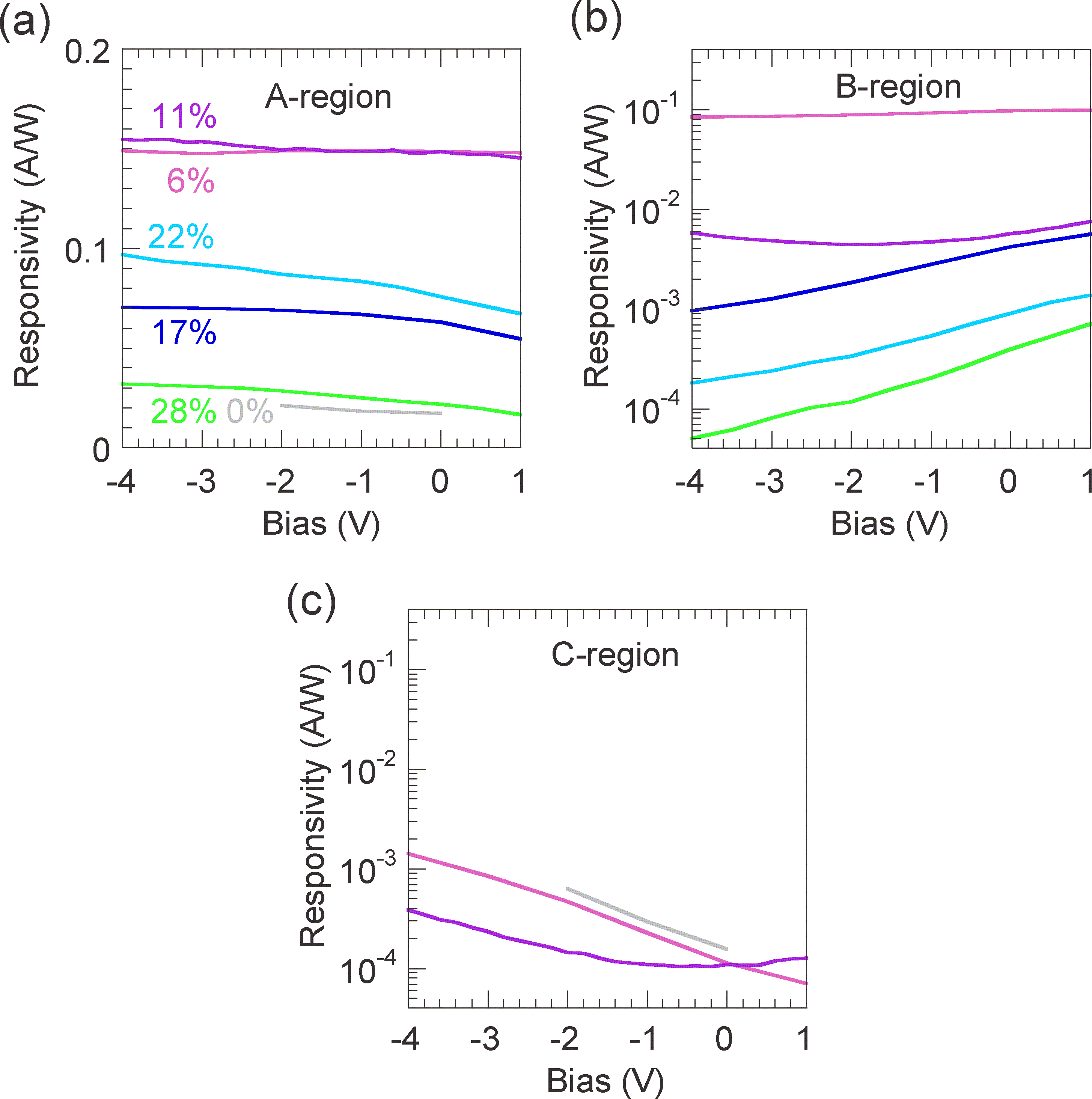}
\caption{(a)-(c) Change in responsivity measured by BPCS as a function of applied bias at exciting photon energies belonging to the A, B, C regions of the spectra for In$_x$Ga$_{1-x}$N QW structures with different indium composition. The photon energies corresponding to the different curves are marked as 'A', 'B', 'C' in Fig.~\ref{fig:responsivity1}.}
\label{fig:responsivity2}
\end{figure}

BPCS measurements were carried out on the series of In$_x$Ga$_{1-x}$N QW structures with indium composition varying between 0\% and 28\%, covering the UV to green range of the spectrum, and for applied biases varying from +1~V to -4~V (where the negative sign indicates a reverse bias). The results are presented in Fig.~\ref{fig:responsivity1}. The responsivity is obtained normalizing the photocurrent measured at a given photon energy to the incident optical power on the sample. Three different regions, named 'A', 'B' and 'C' (enclosed in Fig.~\ref{fig:responsivity1} by boundaries), can be identified in the series of spectra according to their bias dependence. The bias dependence in these three regions is shown more in detail in Fig.~\ref{fig:responsivity2}, where the responsivity at some fixed energy value is recorded for a continuous variation of the applied bias from -4 to +1~V. The specific energy values chosen for each sample are indicated in Fig.~\ref{fig:responsivity1}. The photoluminescence (PL) peak energy of the solar cells measured at a bias of 0~V with different above-gap excitation laser sources is marked as stars in Fig.~\ref{fig:responsivity1}.

The A-region is characterized by a weak increase in responsivity at increasing reverse bias [Fig.~\ref{fig:responsivity2}(a)]. It is observed for all indium compositions (Fig.~\ref{fig:responsivity1}) and corresponds to absorption above the band gap of the QWs, $E_g$. The latter is defined, similarly to Ref.~\citenum{Helmers2013}, as the transition energy between the exponential onset of the Urbach tail and the saturation region. From this definition it can be seen in Fig.~\ref{fig:responsivity1}(b)-(f) that $E_g$ lies close to the intersection of the BPCS curves at the boundary between the A- and B-region.\bibnote{More precisely, the so-defined band gap energy slightly varies due to the blue-shift of the responsivity curves with increasing reverse bias. However its value remains quite close, within 50 meV, to the boundary between the A- and the B-region.} A detailed description of the carrier escape mechanism from the QWs for excitation in the A-region is given in Ref.~\citenum{Lang2012} according to which both tunneling and thermionic emission are expected to contribute, considering a barrier thickness of 7~nm as in our structures. No, or very little, disorder-induced effects are expected for this process, so the A region will not be discussed in the following.

The B-region is identified by an exponential decrease in responsivity at increasing reverse bias [Fig.~\ref{fig:responsivity2}(b)], corresponding in the photocurrent spectra to an Urbach tail undergoing a blue-shift with increased reverse voltage (Fig.~\ref{fig:responsivity1}). Let us recall that in c-plane wurtzite InGaN/GaN structures, the electric field $\vec{E}_{QW}$ points towards the substrate, as shown in the schematic of Fig.~\ref{fig:responsivity1}(f).\cite{Feneberg2007} Increasing the reverse bias then partly compensates $\vec{E}_{QW}$ leading to a blue-shift in the absorption edge. The B-region appears in the spectra for [In]=6\% and its energy range progressively extends at larger indium concentrations. We remark that the PL peak energy of all the In$_x$Ga$_{1-x}$N/GaN solar cells lies in the B-region which then is associated with transitions occurring in the QWs. Below-gap absorption processes giving origin to the B-region may depend on compositional disorder and QW electric field, as it will be further analyzed in Sec.~IV.

None of these effects are expected to significantly affect below-gap absorption in semipolar GaN QWs, due to the reduced internal electric fields and absence of alloy fluctuations, explaining why the B-region is not observed in this sample [absorption curves in Fig.~\ref{fig:responsivity1}(a)]. Let us point out that this result does not exclude the occurrence of compositional disorder effects in semipolar (or non-polar) ternary InGaN layers, which were not investigated in the present work.\textcolor{red}{\cite{Monavarian2016, Shahmohammadi2017}}

Finally, the C-region, in contrast with the B-region, is characterized by an exponential increase in responsivity at increasing reverse bias [Fig.~\ref{fig:responsivity2}(c)] and is most clearly observed in the UV diodes. We interpret it as being due to type~II well-to-barrier transitions, as detailed in the Appendix.

We remark that the B- and C-region originate from competing mechanisms of below-gap absorption which dominate, respectively, at high and low indium concentration. The sample with [In]=11\% corresponds to the intermediate case in which the two mechanisms have comparable efficiencies as may be observed in the deviation from the typical responsivity trend of the B- and C-region at large and small reverse bias, respectively [purple curves in Fig.~\ref{fig:responsivity2}(b and (c)].

In the following the dependence of the Urbach tail observed in the B-region will be examined as a function of different factors, such as bias, indium composition, sample temperature and type of device, specifically solar cells or LEDs both incorporating InGaN QWs.

\begin{figure}
\includegraphics[width=0.49\textwidth]{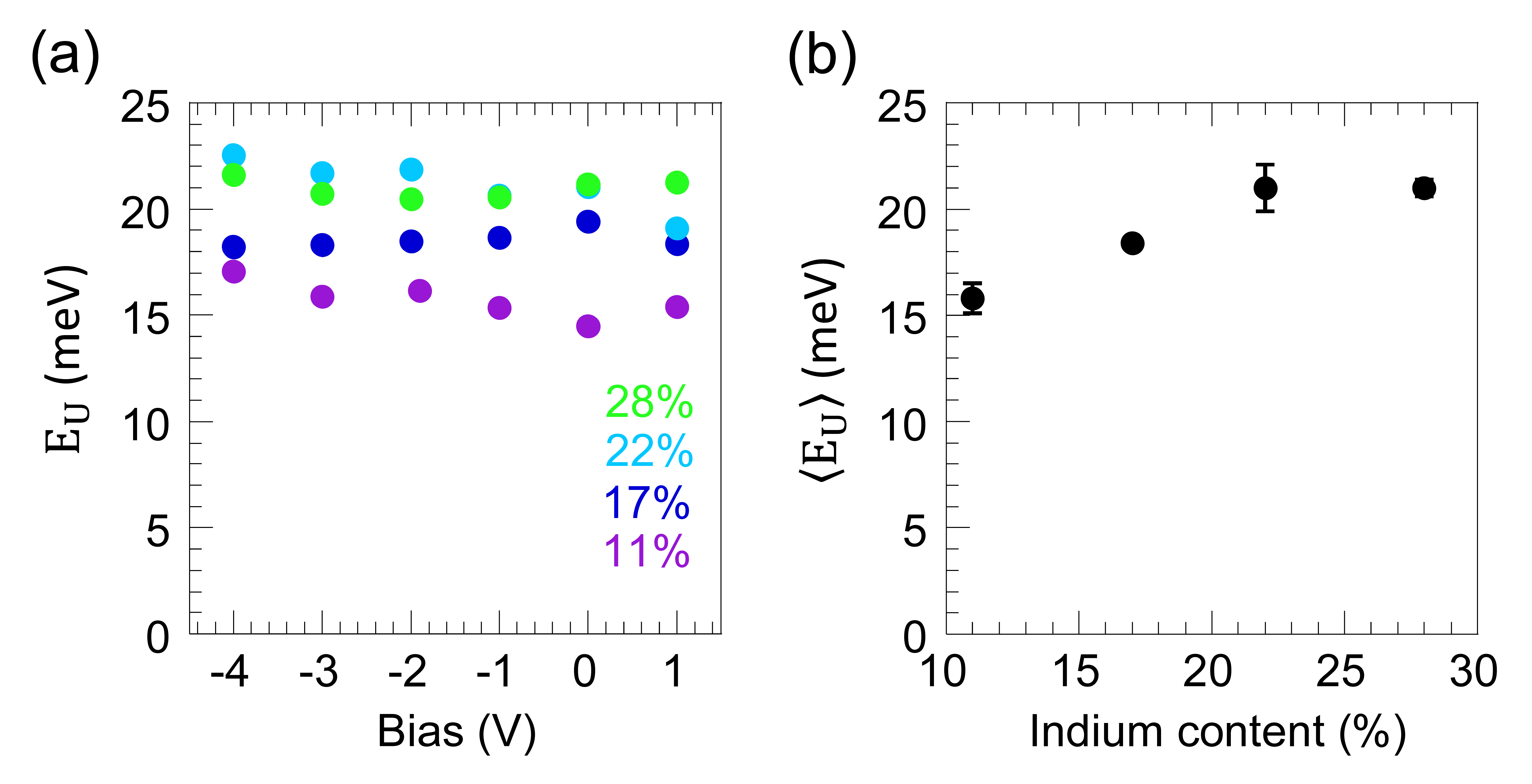}
\caption{(a) Dependence on applied bias of the Urbach energy $E_U$ determined in the B-region of BPCS spectra for In$_x$Ga$_{1-x}$N QW layers with different mean indium content. (b) Mean values of $E_U$ averaged over the applied biases and plotted as a function of the mean indium content of the corresponding InGaN/GaN solar cell.}
\label{fig:Urbach_energy}
\end{figure}

Urbach energies are fitted in a typical interval of photon energies ranging from $E_g-150$~meV to $E_g-20$~meV for samples with mean indium composition varying between 11\% and 28\%. If carrier collection by BPCS were substantially dependent on the energy of the confined states in the QW, then one would expect that for a larger reverse bias applied to the solar cells the extraction efficiency of low- localization energy carriers would increase, while that of high- localization energy carriers would remain roughly constant, thus changing the slope of the measured Urbach tails. In contrast, the experimental fact that the $E_U$ values appear to be almost independent on the applied bias [Fig.~\ref{fig:Urbach_energy}(a)] indicates that carrier collection is only weakly sensitive to the exciting photon energy in the B-region, corroborating our hypothesis of the responsivity spectra as proportional to optical absorption. The average of $E_U$ taken over the different applied biases and denoted as $\langle E_U \rangle$ in Fig.~\ref{fig:Urbach_energy}(b) varies weakly with indium concentration for QWs emitting from the violet to the green range with the typical value being around 20~meV.

\begin{figure}
\includegraphics[width=0.47\textwidth]{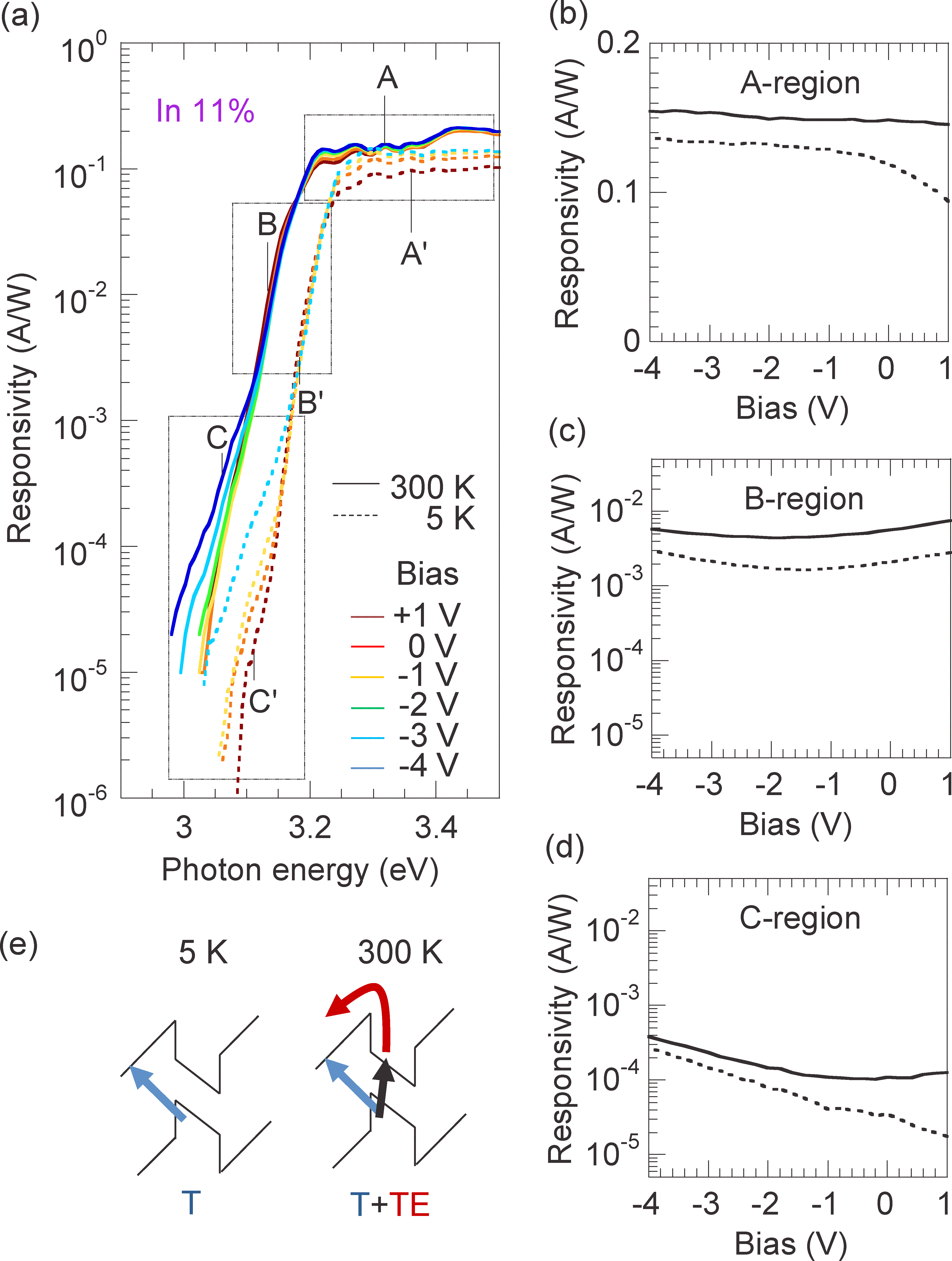}
\caption{(a) Responsivity measurements by BPCS of an In$_{0.11}$Ga$_{0.89}$N/GaN solar cell at 300~K (continuous lines) and 5~K (dashed lines) as a function of applied bias. The low temperature curves are blue-shifted due to the increase in bandgap energy. The A, B, C regions are enclosed by boundaries. (b)-(d) Change in responsivity as a function of applied bias at 300~K (continuous lines) and 5~K (dashed lines) for exciting photon energies belonging to the different regions and marked in (a) as A, B, C and A’, B’, C’, at 300~K and 5~K, respectively. (e) Different mechanisms of carrier extraction from the QW.  At low temperature tunneling outside the QW (T) may occur via type-II transitions. At room temperature, in addition to tunneling, absorption in the QW followed by thermionic emission (TE) may also occur.}
\label{fig:responsivity3}
\end{figure}

In Fig.~\ref{fig:responsivity3}(a) BPCS measurements of the sample with [In] = 11\% at room temperature and at 5~K can be compared. The increase in band gap energy at low temperature causes the photocurrent spectra to shift towards higher photon energy. But the bias dependence of the three different regions remains essentially unaffected [Fig.~\ref{fig:responsivity3}(b)-(d)], except for the C-region in the low-bias range, a feature that is discussed in the Appendix. Remarkably, the slope of the Urbach tail is substantially unchanged by the sample temperature: $\langle E_U \rangle$ is $15.8 \pm 0.7$~meV at 300~K and $14.7 \pm 0.9$~meV at 5~K, in agreement with the interpretation of the Urbach tail as being related with compositional disorder, a structural property of the QWs that is temperature independent,\cite{Sa-Yakanit1987} therefore not due to phonon-assisted broadening.

\begin{figure}
\includegraphics[width=0.48\textwidth]{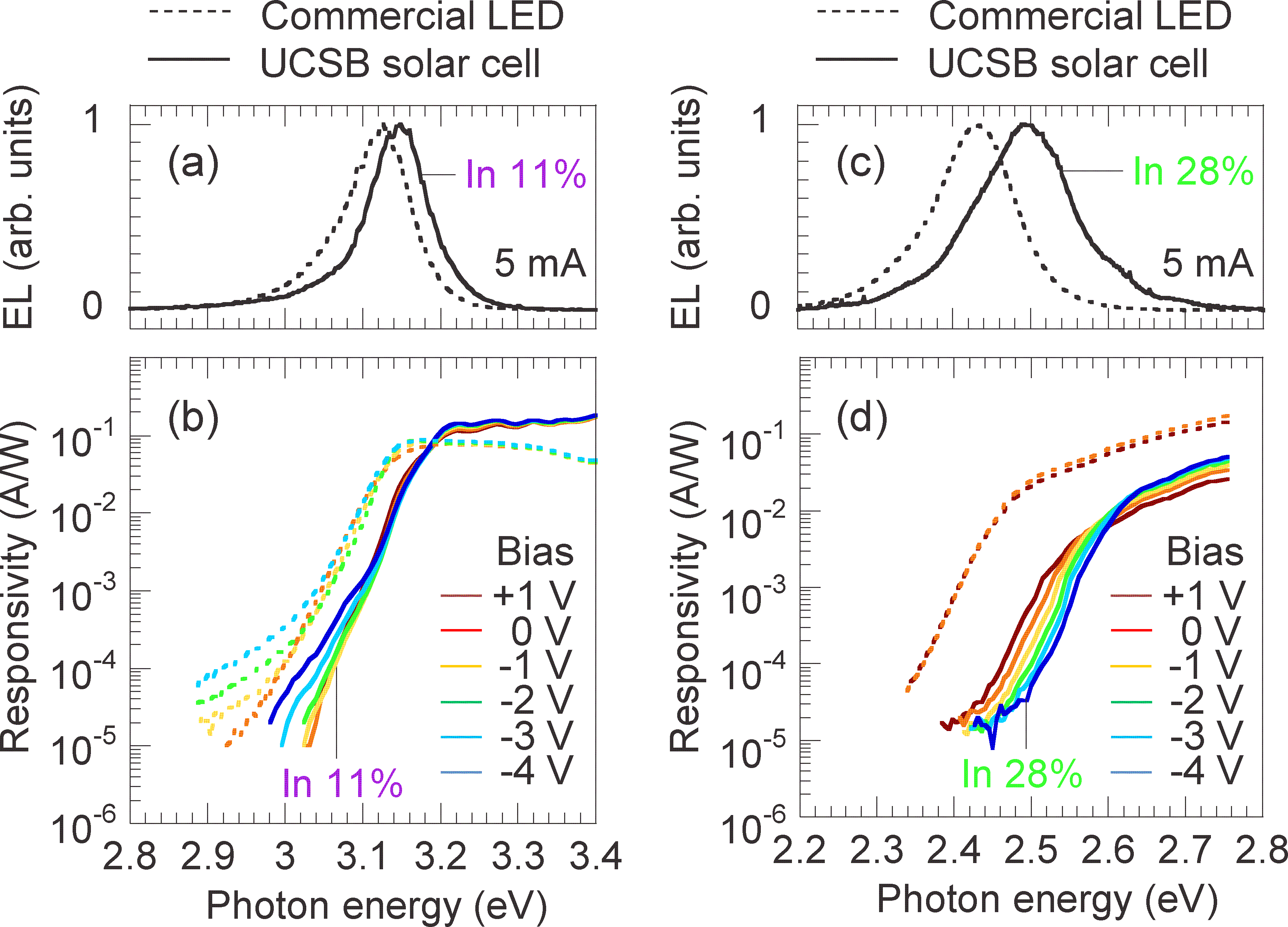}
\caption{(a),(b) EL spectra measured at 5~mA injection and BPCS measurements of an In$_{0.11}$Ga$_{0.89}$N/GaN solar cell (continuous lines) and a commercial UV LED (dashed lines). (c),(d) EL spectra measured at 5~mA injection and BPCS measurements of an In$_{0.28}$Ga$_{0.72}$N/GaN solar cell (continuous lines) and a commercial bluish-green LED (dashed lines).}
\label{fig:ELspectra}
\end{figure}

Next we compare BPCS of UCSB solar cells with commercial LEDs to verify whether commercial-grade wafers and a different optimization of the optoelectronic devices, i.e., luminescence in LEDs vs. carrier collection in solar cells, may give rise to different features in the photocurrent spectra. Two commercial LEDs with electroluminescence (EL) spectra similar to those of two UCSB solar cells with [In]=11\% and 28\% are investigated [Fig.~\ref{fig:ELspectra}(a),(c)]. The corresponding sets of BPCS measurements exhibit strong similarities, apart from a small energy shift likely due to the difference in QW composition. In particular, in the case of UV devices [Fig.~\ref{fig:ELspectra}(b)] the same A, B, C regions characterized by their bias dependence, as previously discussed, are observed. Green-emitting devices also exhibit common features related to the A and B regions [Fig.~\ref{fig:ELspectra}(d)], though the application of a negative bias to the commercial green LED is severely limited by a Zener diode in parallel with the diode. The Urbach energies of the commercial devices are $21.4 \pm 1.1$~meV and $20.5 \pm 0.5$~meV for the UV and green LED, respectively, being closely in the range of $E_U$ measured in the UCSB solar cells. This indicates that $E_U$ is essentially unaffected by any possible difference in material quality. The overall close response in the measured spectra of the different devices suggests that BPCS allows carrying out a spectroscopy of intrinsic optoelectronic properties of InGaN QWs.

\section{Theoretical model}
\label{sec:theor_model}

In this section the possible absorption mechanisms giving origin to the Urbach tail of the B-region are analyzed by means of different models. Considering the strong polarization-induced electric fields present in c-plane InGaN/GaN QWs, typically of the order of few MV/cm in the range of [In] considered here, the first hypothesis we formulate is that the measured Urbach tails originate from FK, a well-known mechanism of below-gap absorption in bulk semiconductors. In presence of an electric field $\vec{E}_{QW}$ the electron and hole wave functions, which are Airy functions with tails extending in the classically forbidden region, overlap even for transition energies $h \nu < E_g$.

\begin{figure}
\includegraphics[width=0.45\textwidth]{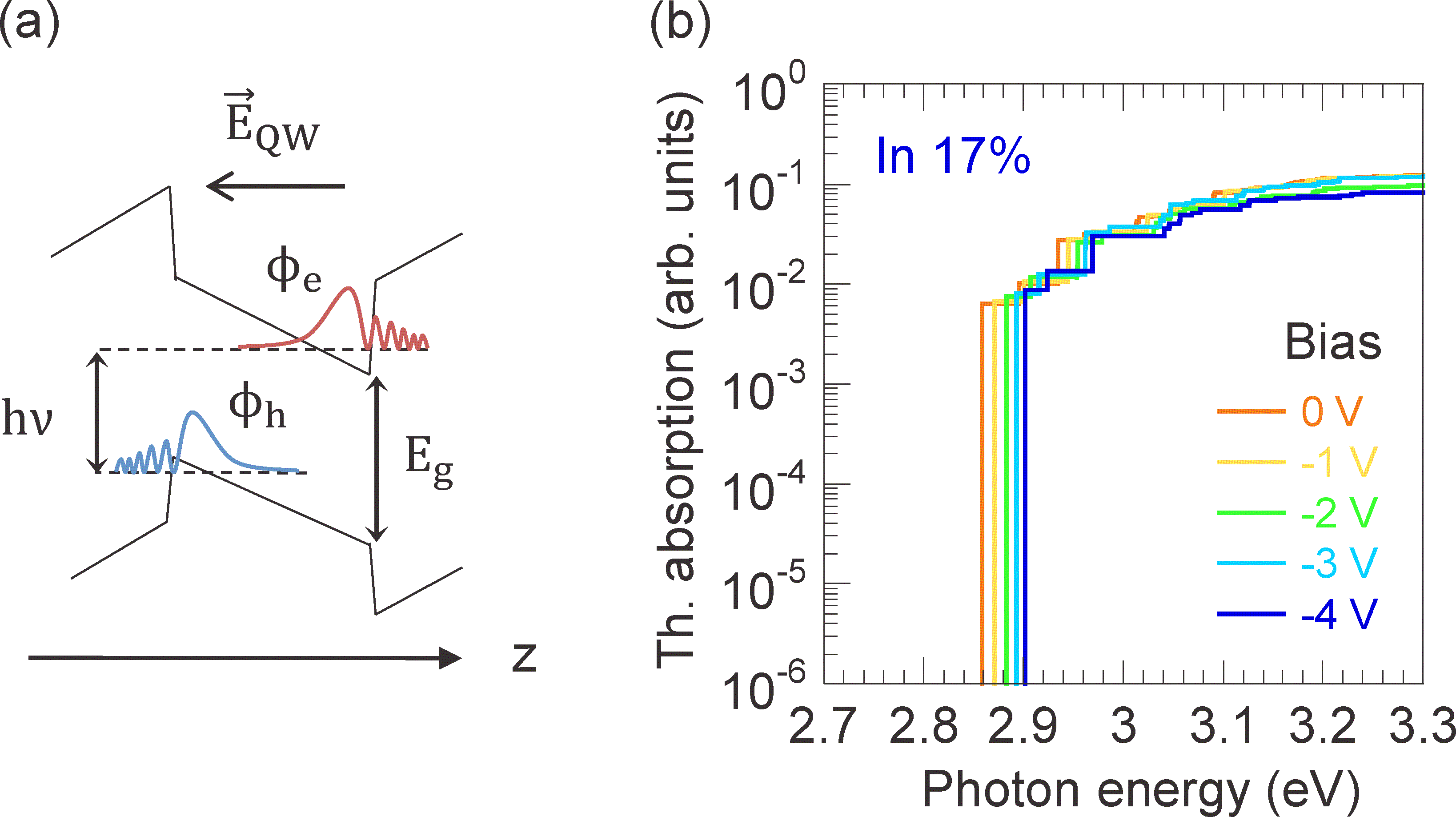}
\caption{(a) Schematic of QCFK absorption in a QW. In presence of an electric field $\vec{E}_{QW}$ below-gap absorption at energy $h\nu < E_g$ is allowed by the overlap of the electron and hole wave functions. (b) Simulation of QCFK absorption curves as a function of applied bias for an In$_{0.17}$Ga$_{0.83}$N/GaN MQW solar cell structure including thickness fluctuations among the different QWs. Alloy disorder is not taken into account in these simulations.}
\label{fig:schematic_absorption}
\end{figure}

Since we are dealing with QW structures, we should consider the quantum-confined Franz-Keldysh effect (QCFK),\cite{Miller1986} shown schematically in Fig.~\ref{fig:schematic_absorption}(a), which corresponds to the FK in the limit of thin slabs of material. The absorption curve of an ideal QW should exhibit a step-like increase with photon energy due to the discrete nature of the allowed interband transitions. As a consequence of the symmetry breaking caused by the electric field, the major effect of QCFK is to change the selection rules in the QW and permit ``forbidden'' transitions.

To evaluate the influence of $\vec{E}_{QW}$ in below-gap absorption we simulate the absorption curves of the In$_{0.17}$Ga$_{0.83}$N/GaN MQW solar cell structure as a function of bias using a 1D~Schr\"odinger-Poisson-drift-diffusion solver~\cite{Wu2015b, deFalco2005} that assumes \textit{homogeneous} QWs, i.e., without alloy disorder, but including thickness fluctuations of $\pm 1$~monolayer among the different QWs.\bibnote{The thickness fluctuations among the different QWs are assumed to follow approximately a Gaussian distribution, namely we consider 5~QWs being 1.8~nm thick (XRD value for [In]=17\%), 2~QWs 1.5~nm thick and 3~QWs 2.1~nm thick.} The polarization fields are calculated from the spontaneous and piezoelectric polarizations of nitride QWs (parameters given in Ref.~\citenum{Li2017}). The quantum mechanical expression of the absorption coefficient of Ref.~\citenum{Miller1986} is used, taking into account both transitions from the heavy hole and light hole bands. The simulated absorption spectra are shown in Fig.~\ref{fig:schematic_absorption}(b). The blue-shift of the absorption edge at increasing reverse bias due to the partial compensation of $\vec{E}_{QW}$ is quantitatively well reproduced in the simulations. Nevertheless the simulated curves exhibit a step-like increase due to the small thickness of the QW layers~\cite{Miller1986} and QCFK alone is not sufficient to explain the observed experimental broadening of the absorption edge [cf. Fig.~\ref{fig:responsivity1}(d)]. Based on low temperature measurements discussed in Sec.~III phonon-assisted broadening of the absorption edge has already been excluded. The main remaining mechanism of broadening is compositional disorder. Therefore to properly describe the Urbach tails of InGaN layers we develop a 3D absorption model incorporating both the effects of the electric field and compositional disorder in the QW.

In principle, to compute the absorption coefficient of a QW, solving the Schr\"odinger equation is necessary to obtain the localized eigenstates and corresponding eigenenergies needed to calculate the electron-hole overlap and energies of the different interband optical transitions.\cite{Singh2007} In addition to this, when simulating biased devices, such as the solar cells considered in this work, also the Poisson and drift-diffusion (DD) equations have to be solved to account for the charge distribution and carrier transport. Note that solving the coupled Schr\"odinger-Poisson-DD equations for a 3D~device structure incorporating compositional disorder can be very demanding in terms of computational resources (see LL3, Ref.~\citenum{Li2017}). Here we employ an alternative approach based on the LL~theory\cite{Filoche2012} which is able to capture the confining properties of a disordered system without solving the associated Schr\"odinger equation
\begin{equation}\label{eq:Schrodinger}
\widehat{H} \psi = E~\psi~,
\end{equation}
where $\widehat{H} = -\hbar^2/2m~\Delta + V$ is the Hamiltonian, $V$ is the potential energy of the disordered system and $\psi$ and $E$ are the eigenstate and eigenenergy, respectively. Instead, a much simpler linear problem defined by
\begin{equation}\label{eq:landscape}
\widehat{H} u = 1
\end{equation}
is solved. Arnold et al. recently demonstrated that the inverse of this landscape, i.e. $1/u$, acts as an ``effective'' confining potential seen by the localized eigenstates.\cite{Arnold2016} In other words, the quantum properties of a disordered potential such as quantum wave interference, quantum confinement and tunneling are directly translated into $1/u$ in the form of a classical potential. An important property of the function $1/u$ (or $u$) is that its crest (respectively, valley) lines partition the domain into the localization subregions, i.e., the regions of lower energy solutions of the Schr\"odinger equation (LL1, Ref.~\citenum{Filoche2017}). This allows a very efficient computation of quantum effects in disordered semiconductors, and will be used in the following to compute the absorption spectra of the measured nitride solar cells.

\begin{figure}
\includegraphics[width=0.47\textwidth]{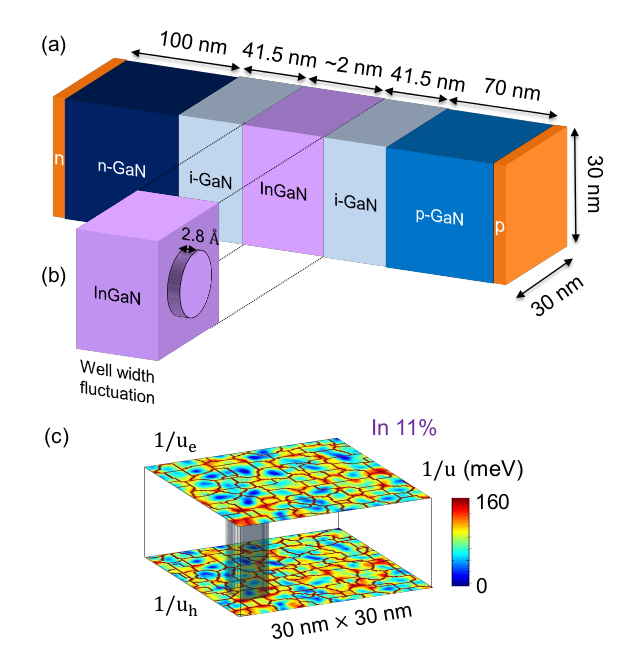}
\caption{(a) Schematic of the typical InGaN/GaN QW structure used in the 3D~absorption model. (b) The simulated InGaN QW incorporates a disk that is one monolayer thick representing a well width fluctuation. (c) ``Effective'' electron and hole confining potentials obtained from the LL~theory for an In$_{0.11}$Ga$_{0.89}$N/GaN structure at 0~V and computed self-consistently with the Poisson-DD equations. The maps shown here correspond to 2D~cuts of the 3D~landscapes in the middle plane of the QW.}
\label{fig:schematic_structure}
\end{figure}

A schematic of the typical simulated InGaN/GaN QW structure is shown in Fig.~\ref{fig:schematic_structure}(a). The size of the simulated domain is 30~nm $\times$ 30~nm $\times$ 255~nm. For the ease of computation the MQW region of the measured InGaN/GaN solar cells is substituted by an equivalent intrinsic layer with a single QW. Since the 10~QWs of each solar cell are nominally identical with very similar electric field conditions, the single QW~model represents a fair approximation of the MQW structure in terms of the absorption spectrum. For each solar cell, the simulations are repeated for 10~different random configurations of the indium map and finally averaged to produce an average absorption curve. The indium atoms are first randomly distributed in the QW on an atomic grid with spacing a=2.833~\AA{ }corresponding to the average distance between cations in GaN. Then at each grid site the local indium composition is averaged via the Gaussian averaging method (see LL3, Ref.~\citenum{Li2017}), allowing to smooth the rapidly oscillating distribution of atoms and to obtain a continuous fluctuating potential. In order to preserve the disorder effects on the electronic properties of the QW, the length scale of the averaging must be smaller or comparable to the typical scale of the effective potential fluctuations seen by the carriers. We fixed the Gaussian broadening parameter to a value of 2.0a$\approx$0.6 nm, which satisfies the aforementioned condition (LL3, Ref.~\citenum{Li2017}). By taking the plane-averaged values of the resulting indium composition we obtain an indium profile along the growth direction, whose maximum defines the indium concentration and whose FWHM defines the thickness of the QW. In the following simulations we model four solar cell structures with InGaN QWs having an indium composition of 12\%, 18\%, 22\%, 28\% and a corresponding FWHM of 1.6 nm, 1.8 nm, 1.8 nm and 2.1 nm. These values allow to align the threshold of the simulated absorption spectra with that of the experimental curves, while remaining within $\sim$1\% of the indium content and 0.1~nm of the QW thickness values determined by XRD in the experimental structures. Moreover, to include in the simulations the fluctuations in the width of the QWs, we consider in the process of random indium atoms generation a circular region with one additional monolayer [Fig.~\ref{fig:schematic_structure}(b)], similarly to Ref.~\citenum{Schulz2015}. The disk occupies roughly 1/3 of the entire QW~area.

All the material parameters needed in the simulations are assigned according to the local composition by an interpolation method as
\begin{equation}
\begin{aligned}
E_g^{\mbox{In}_x\mbox{Ga}_{1-x}\mbox{N}} = &~ (1-x)~E_g^{\mbox{GaN}} + x~E_g^{\mbox{InN}} \\& \quad - 1.4~x(1-x),\\
E_g^{\mbox{Al}_x\mbox{Ga}_{1-x}\mbox{N}} = &~ (1-x)~E_g^{\mbox{GaN}} + x~E_g^{\mbox{AlN}} \\& \quad - 0.8~x(1-x),\\
\varepsilon_r^{\mbox{In}_x\mbox{Ga}_{1-x}\mbox{N}} = &~(1-x)~\varepsilon_r^{\mbox{GaN}} + x~\varepsilon_r^{\mbox{InN}},\\
\varepsilon_r^{\mbox{Al}_x\mbox{Ga}_{1-x}\mbox{N}} = &~(1-x)~\varepsilon_r^{\mbox{GaN}} + x~\varepsilon_r^{\mbox{AlN}},\\
m^{*,\mbox{In}_x\mbox{Ga}_{1-x}\mbox{N}} = &~\left((1-x)/m^{*,\mbox{GaN}} + x/m^{*,\mbox{InN}}\right)^{-1},\\
m^{*,\mbox{Al}_x\mbox{Ga}_{1-x}\mbox{N}} = &~\left((1-x)/m^{*,\mbox{GaN}} + x/m^{*,\mbox{AlN}}\right)^{-1}
\end{aligned}
\end{equation}
where the GaN, AlGaN and InGaN parameters are given in Table~\ref{tab:parameter-band}. The new self-consistent Poisson-DD-landscape approach presented in LL1 (Ref.~\citenum{Filoche2017}) is used to solve the Poisson and DD equations self-consistently with Eq.~(\ref{eq:landscape}) accounting for quantum effects of disorder. Before solving the Poisson equation, the spontaneous and piezoelectric polarization fields are calculated over the entire structure, as detailed in LL3 (Ref.~\citenum{Li2017}). For each given bias applied to the simulated solar cell the 3D~LLs are computed. As an example, 2D cuts of the ``effective'' confining potentials of electrons and holes, $1/u_e$ and $1/u_h$, in the middle of the QW of an In$_{0.11}$Ga$_{0.89}$N/GaN structure at 0~V are shown in Fig.~\ref{fig:schematic_structure}(c), where the network of localization subregions is highlighted (thick lines). We shall now see how to exploit the LLs to compute absorption in these structures.

\begin{table}[tb!]
\begin{center}
\begin{tabular}{ccccccc}
\hline 
\hline
& $E_g$  & $\varepsilon_r$ & $m_{e}^{\parallel}$ & $m_{e}^{\perp}$ & $m_{hh}$ & $m_{lh}$ \\
~units~ & (eV) &  & ($m_0$) & ($m_0$) & ($m_0$) & ($m_0$)
\\ \hline
~GaN~  & 3.437 & 10.4  & 0.21 & 0.20 & 1.87 & 0.14 \\
~InN~  & 0.61  & 15.3  & 0.07 & 0.07 & 1.61 & 0.11 \\
~AlN~  & 6.0   & 10.31 & 0.32 & 0.30 & 2.68 & 0.26 \\
\hline
\multicolumn{4}{c}{Bandgap alloy} & \multicolumn{3}{c}{InGaN: 1.4} \\
\multicolumn{4}{c}{bowing parameter} & \multicolumn{3}{c}{AlGaN: 0.8} \\
\hline
\hline
\end{tabular}
\end{center}
\caption{Band structure parameters for wurtzite nitride alloys: bandgap, relative permittivity, and effective masses.\cite{Vurgaftman2001, Piprek2007}}
\label{tab:parameter-band}
\end{table}

In the companion paper LL1 (Ref.~\citenum{Filoche2017}), the LL~theory is applied to the framework of semiconductors to derive a model accounting for quantum effects in disordered materials. In particular it is shown that the landscape $u$ allows estimating in each localization subregion $\Omega_m$ the fundamental localized eigenfunction and its corresponding eigenenergy as:
\begin{align}
\psi_0^{(m)} &\approx \frac{u}{\|u\|} \label{eq:psi0}\\
E_0^{(m)} &\approx \frac{\braket{u|1}}{\|u\|^2} \label{eq:E0}
\end{align}
where the expressions above must be evaluated in the considered subregion and the L$^2$-normalization of $u$ is used. Moreover the following expression for the absorption coefficient of a disordered QW is derived
\begin{equation}\label{eq:absorption_QW_landscape}
\alpha(hv) = \frac{2}{3}\cdot C \cdot {\rm JDOS}_{3D}^{(m,n)}(hv)~I_{0,0}^{(m,n)}
\end{equation}
where $C=\pi e^2 \hbar |p_{cv}|^2/m_0^2 n_r \varepsilon_0 h \nu$ is a prefactor depending on the real part of the surrounding refractive index $n_r$ and the interband momentum matrix element $p_{cv}$, and where the summation is carried out over all electron and hole subregions (labeled $m$ and $n$, respectively). Superimposing the maps of the two landscapes [Fig.~\ref{fig:schematic_structure}(c)], several subregions which consist in intersections of the localization subregions of electrons and holes can be defined. The joint density of states of each of these electron-hole ``overlapping'' subregions is given by
\begin{align}\label{eq:JDOS}
{\rm JDOS}_{3D}^{(m,n)}\left(h\nu\right) &= \nonumber\\
\frac{\sqrt{2}m_r^{3/2}}{\pi^2 \hbar^3}~&\sqrt{h\nu - E_g - E_{e,0}^{(m)} - E_{h,0}^{(n)}} \quad,
\end{align}
where $m_r=\left(1/m_e + 1/m_h\right)^{-1}$ is the reduced effective mass, and $E_{e,0}^{(m)}$ and $E_{h,0}^{(n)}$ are the fundamental localized states of the $m$-th electron and $n$-th hole subregions. The overlap factor is determined by the electron and hole fundamental eigenstates as
\begin{equation}\label{eq:I00}
I_{0,0}^{(m,n)} = \frac{|\braket{\psi_{e,0}^{(m)}|\psi_{h,0}^{(n)}}|^2}{\|\psi_{e,0}^{(m)}\|^2 \|\psi_{h,0}^{(n)}\|^2}
\end{equation}
Note that the eigenenergies and eigenfunctions appearing in Eq.~(\ref{eq:JDOS}) and (\ref{eq:I00}) can be directly computed from the LLs of electrons and holes using Eq.~(\ref{eq:psi0}) and (\ref{eq:E0}), respectively, without having to resort to the Schr\"odinger equation. In support of this absorption model, we have shown in Sec.~IV~B of LL1 (Ref.~\citenum{Filoche2017}) that the absorption spectrum calculated in a 1D disordered superlattices by the landscape theory agrees very well with the exact 1D Schr\"odinger calculation.

\begin{figure}
\includegraphics[width=0.49\textwidth]{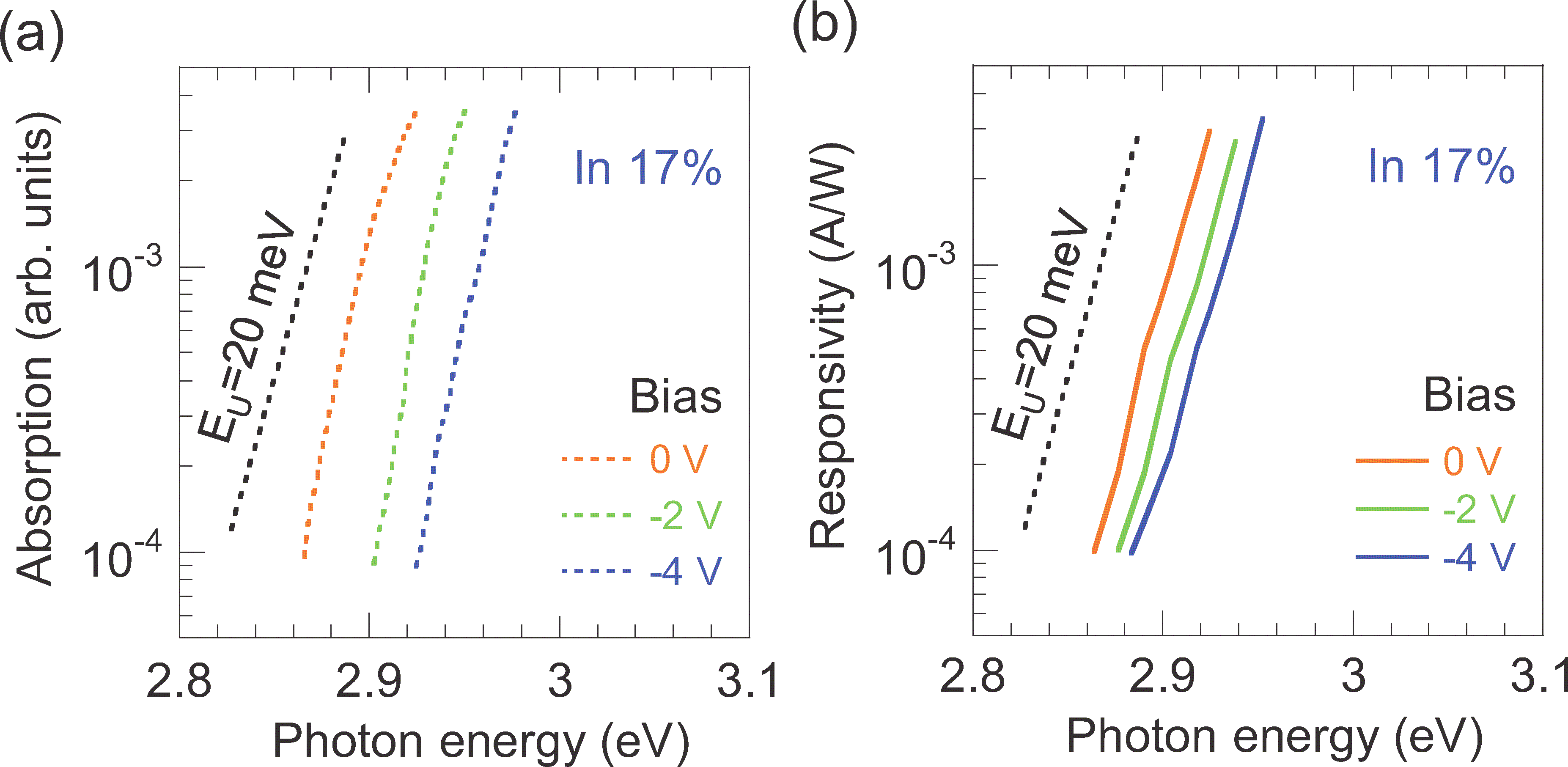}
\caption{Urbach tails of In$_{0.17}$Ga$_{0.83}$N/GaN structures: (a) computed from the 3D~absorption model including compositional disorder and QW electric field, and (b) experimentally measured by BPCS. Curves correspond to different applied biases to the structure. The dashed line is shown as a reference and corresponds to an Urbach energy of 20~meV.}
\label{fig:Urbach_tails}
\end{figure}

\begin{figure}
\includegraphics[width=0.42\textwidth]{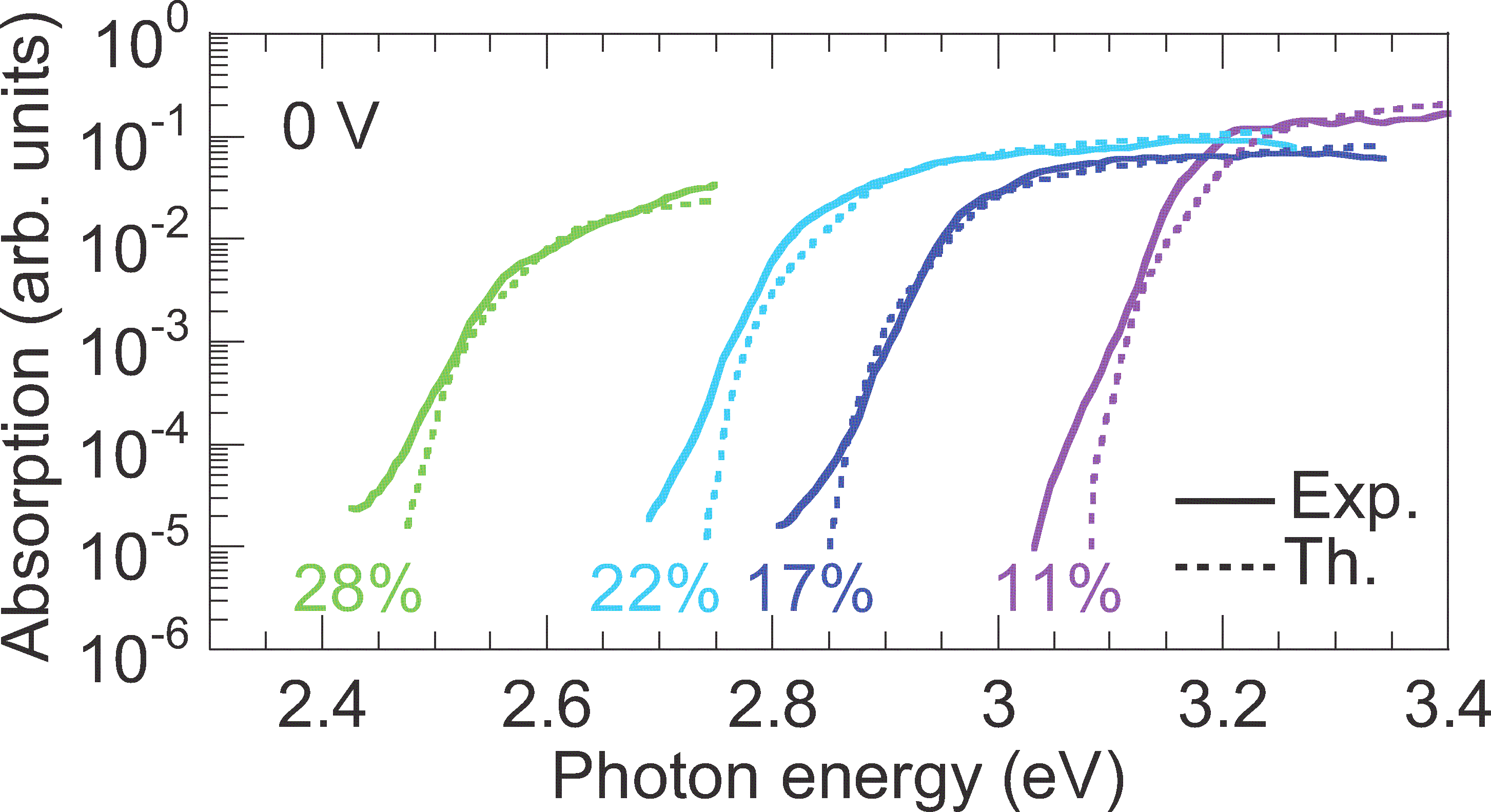}
\caption{Absorption curves for InGaN/GaN structures of different indium composition at 0~V computed using the 3D~theoretical model incorporating compositional disorder and QW electric field as a function of different smoothing parameters (dashed line). The responsivity curves measured by BPCS of InGaN/GaN solar cells of different indium composition at 0~V are also shown (continuous lines).}
\label{fig:absorption_curves}
\end{figure}

Urbach tails computed from the 3D absorption model for an In$_{0.17}$Ga$_{0.83}$N/GaN structure at different applied biases are shown in Fig.~\ref{fig:Urbach_tails}(a) and can be compared with the experimental Urbach tails measured by BPCS of the In$_{0.17}$Ga$_{0.83}$N/GaN solar cell in Fig.~\ref{fig:Urbach_tails}(b). The 3D~nature of the model allows reproducing both the broadening and the bias-dependence of the experimental Urbach tails which are mainly defined by the in-plane compositional disorder and the QW electric field along the growth direction, respectively. The discrepancy in the magnitude of the blue-shift, differing by a factor $\sim$2 between theory and experiments, may be due to a difference in the fraction of reverse voltage dropped across the MQW active region of the experimental structures and the single QW embedded in an equivalent intrinsic layer used in the simulations. However it should be noted that a quantitative description of the bias-dependence is secondary in terms of the present study, which aims at determining the disorder-induced broadening of the absorption edge, and the observed qualitative agreement is already a good indication that the LL~model includes as well QCFK effects. In Fig.~\ref{fig:absorption_curves} the simulated absorption curves of the InGaN/GaN solar cells at 0~V for a mean indium composition ranging between 11\% and 28\% can be compared with the corresponding experimental measurements. Note that the absorption model only considers optical transitions occurring within the QW which give origin to the A- and B-region observed in the experiments, while type-II well-to-barrier transitions responsible for the C-region and unrelated with disorder (see Appendix) are not taken into account. The simulations reproduce well the absolute energy position and the broadening of the absorption edge as a function of the mean indium content of the QWs. Let us remark that the absorption model incorporating compositional alloy disorder, well width fluctuations and electric field effects gives a lower bound on the Urbach energies, explaining why the simulated tails slightly underestimate the experimental measurements. There may be other broadening processes (e.g. defect-induced smearing of band edges) which could decrease the slope of the below-gap absorption tail (thus increase $E_U$) but these do not seem to play an important role as the agreement with the experiments appears to be already good.

\begin{figure}
\includegraphics[width=0.49\textwidth]{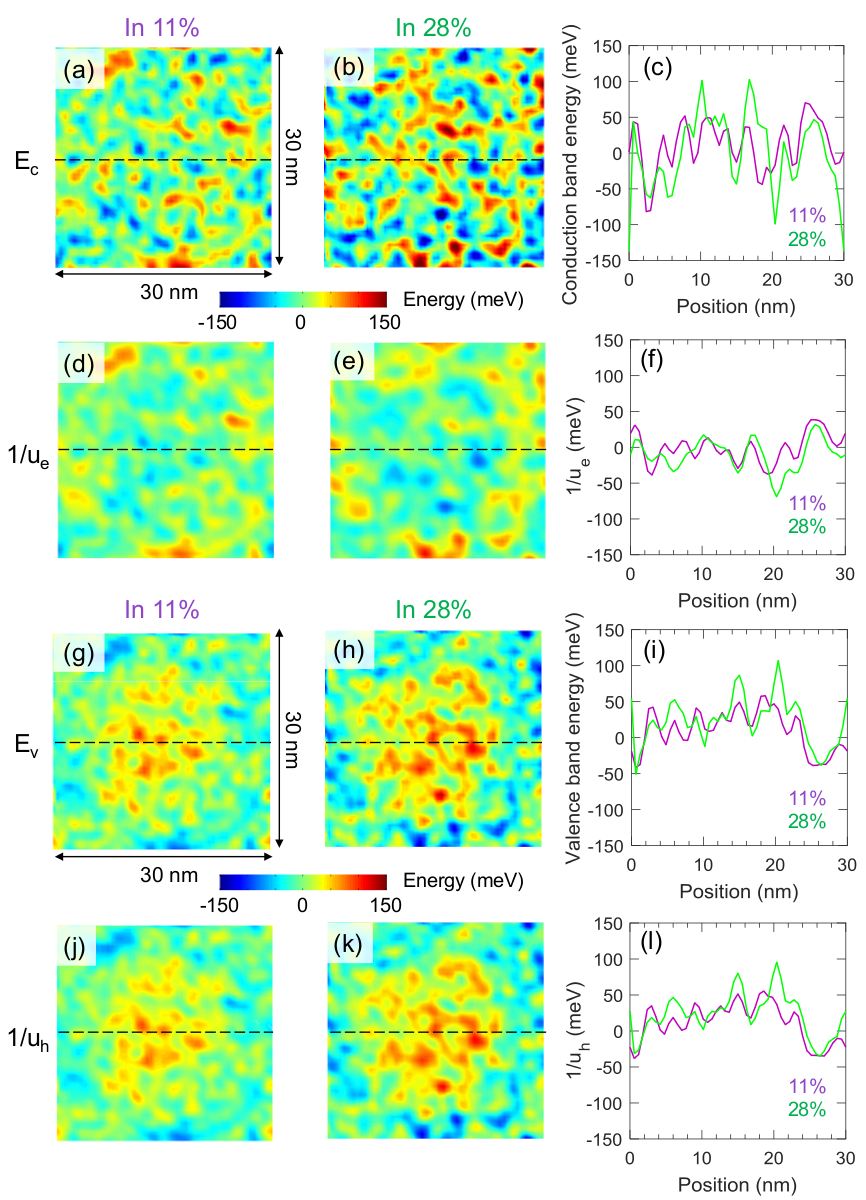}
\caption{(a)-(b)~Simulated conduction band potential maps of InGaN QWs for two different indium compositions (11\% and 28\%). All 2D~cuts in this figure are performed just above $z=141.5$~nm (see Fig.~\ref{fig:schematic_structure}), i.e., at the bottom of the QW. (c)~1D~cuts along the dashed lines of the $E_c$ maps showing the fluctuating potential. (d)-(e)~Electron effective confining potentials calculated from the LL~theory based on the $E_c$ maps shown in (a)-(b). (f)~1D~cuts along the dashed lines of $1/u_e$. The amplitude of the fluctuations is much reduced with respect to that of the corresponding $E_c$  potentials. The simulated maps and 1D~profiles corresponding to the case of holes are shown in (g)-(l). The circular shaped fluctuation observed in the center of (a),(b),(d),(e),(g),(h),(j),(k) is due to the single monolayer disk fluctuation (see Fig.~\ref{fig:schematic_structure}).}
\label{fig:potential_maps}
\end{figure}

\begin{figure}
\includegraphics[width=0.48\textwidth]{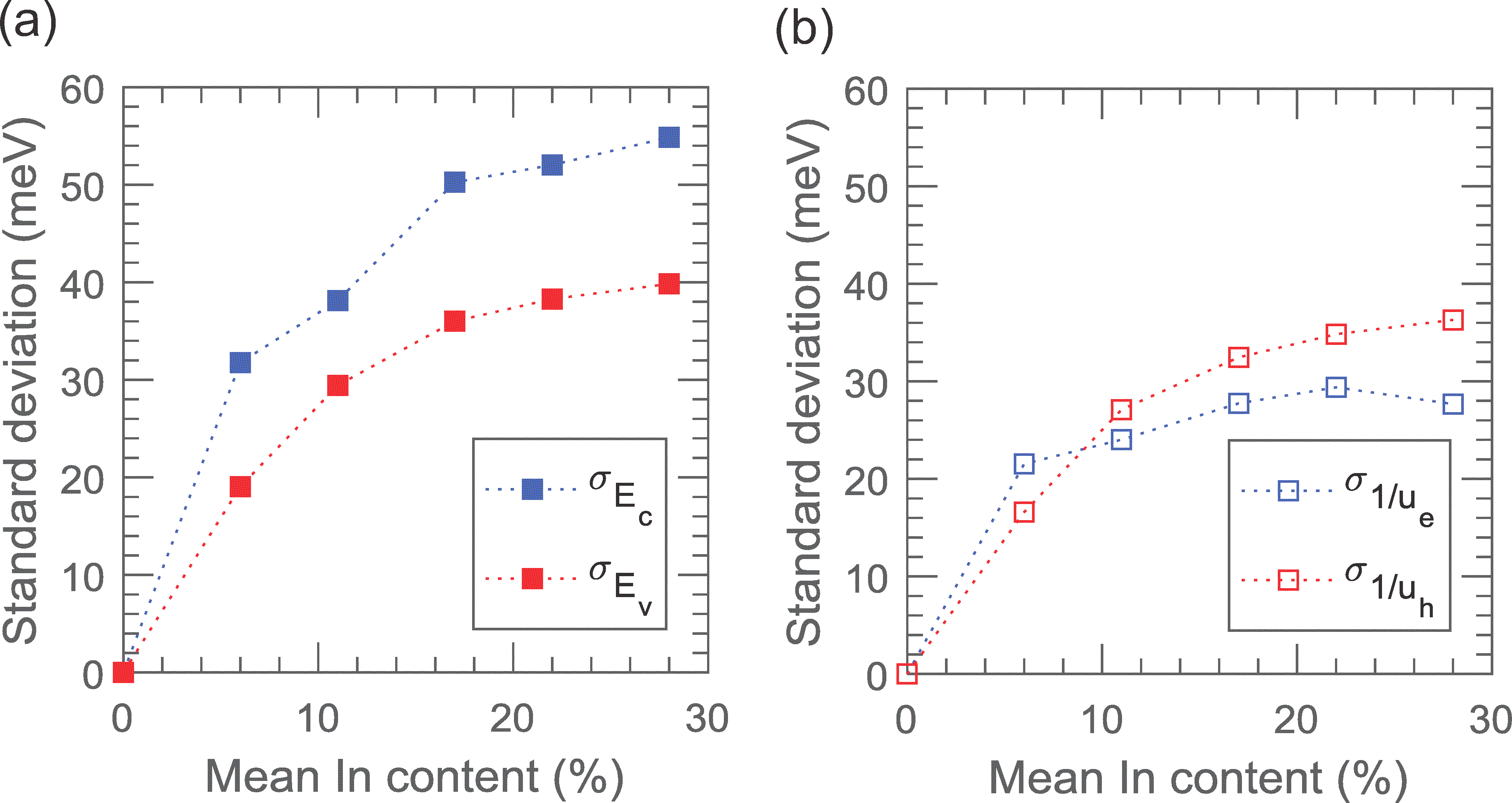}
\caption{Standard deviations characterizing the amplitude of the fluctuations in InGaN QWs of different indium composition for (a) the conduction and valence band potentials, and (b) the corresponding electron and hole effective confining potentials.}
\label{fig:std_dev}
\end{figure}

Finally let us detail why the observed $E_U$ values of $\sim$20~meV [Fig.~\ref{fig:Urbach_energy}(b)] may be surprising and the agreement between the theory and the experiments is remarkable. In the considered range of indium concentrations, compositional disorder is expected to produce large potential fluctuations in the InGaN/GaN QWs. This is clearly observed in the simulations of the conduction band potential $E_c$ for [In]=11\% and [In]=28\% shown in Fig.~\ref{fig:potential_maps}(a)-(c), where the 2D~maps correspond to the mid-plane of the QW and the oscillating curves correspond to 1D~cuts of these maps. The amplitude of the $E_c$ fluctuations, characterized by the standard deviation $\sigma_V$, ranges between 38~meV ([In]=11\%) and 55~meV ([In]=28\%). Intuitively, one could think that the typical size of the potential energy fluctuations of the QW defined by $\sigma_V$ should be comparable to the broadening parameter $E_U$ of the below-gap absorption tails, both being directly related to the DOS of the system. However the $E_U$ values measured in the InGaN  solar cells are quite smaller than $\sigma_V$. Our model based on the LL~theory allows explaining this discrepancy. Indeed by computing the effective potentials $1/u_e$ [Fig.~\ref{fig:potential_maps}(d)-(f)] corresponding to the conduction band maps [Fig.~\ref{fig:potential_maps}(a)-(c)] of the InGaN QWs it can be observed that the amplitude of the fluctuations of the ``effective'' potential, characterized by the standard deviation $\sigma_{1/u}$, is much reduced with respect to $\sigma_V$. The corresponding simulations for the case of holes show a smaller change between the amplitude of the fluctuations of $E_v$ [Fig.~\ref{fig:potential_maps}(g)-(i)] and $1/u_h$ [Fig.~\ref{fig:potential_maps}(j)-(l)], due to the fact that holes have less confinement energy than electrons, therefore are more localized and thus more sensitive to the modulating potential. This is an indication of stronger hole localization than electron one, a phenomenon predicted by different models in the literature.\cite{Schulz2015, WatsonParris2011} For both electrons and holes the values of $\sigma_{1/u}$ lie in the 25-35~meV range, being closer than $\sigma_V$ to the experimental $E_U$ values. The quantum effects of confinement and tunneling, which diminish the extreme values of the potential fluctuations, have been taken into account by the wave equation~Eq.~(\ref{eq:Schrodinger}) and thus diminish the effective energy fluctuations. This result is extended in Fig.~\ref{fig:std_dev} showing the dependence of $\sigma_V$ and $\sigma_{1/u}$ on different indium concentrations for both the electron and hole landscapes. These observations suggest that the model based on $1/u$ finely captures the effective confining potential seen by the carriers in the disordered InGaN layers, which ultimately allows correctly computing the broadening of the absorption edge of the InGaN/GaN~QWs.

\section{Conclusion}

In this work we investigated, both with experiments and theory, the effect of compositional disorder in the Urbach tails of InGaN layers with different indium composition. Experimentally, Urbach tails can be measured by means of biased photocurrent spectroscopy of InGaN/GaN solar cells and LEDs. The different devices produce very similar spectra indicating that a spectroscopy of intrinsic optoelectronic properties of InGaN QWs can be carried out by BPCS. A detailed analysis of the measured responsivity spectra is required in order to identify different below-gap absorption mechanisms. Below-gap absorption due to type-II well-to-barrier transitions, particularly efficient in near-UV emitting QWs, was observed and modeled. A different mechanism of below-gap absorption corresponding to transitions within the QW was observed to be responsible for the broadening of the absorption edge of InGaN QWs with [In]$~>~$6\%. A 1D~model only considering the effect of QW electric field on absorption due to QCFK allows to quantitatively reproduce the blue-shift seen in the experiments but does not account for the experimental broadening of the absorption edge. Thermal disorder is excluded since the experimental broadening of the Urbach tails is essentially unchanged in measurements at 5~K. A 3D~absorption model based on the LL~theory is used to efficiently compute absorption spectra of InGaN/GaN QW structures without having to resort to the Schr\"odinger equation. The model takes into account compositional disorder, well width fluctuations and electric field effects in the QW and this is sufficient to reproduce quite closely the broadening and the bias-dependence of the experimental Urbach tails. The reason for the agreement between experiments and theory lies in the ability of the LL~theory to describe the effective potentials seen by the carriers in the disordered InGaN layers.

\appendix*
\section{Type-II well-to-barrier transitions}

\begin{figure}
\includegraphics[width=0.47\textwidth]{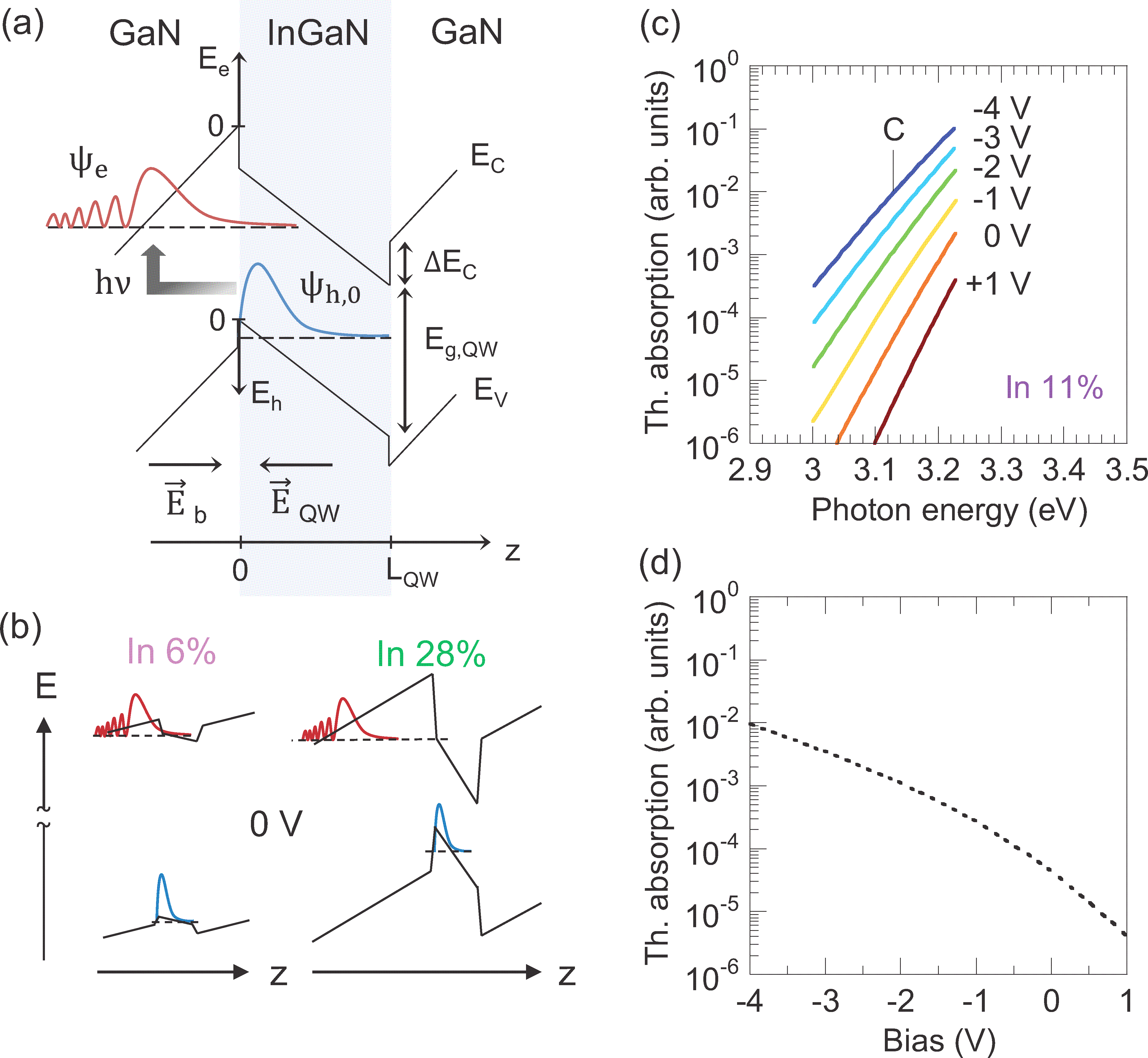}
\caption{(a)~Schematic of type-II well-to-barrier transitions in an InGaN/GaN QW. In presence of barrier and QW electric fields the electron and hole final states, $\psi_e$ and $\psi_{h,0}$, are Airy functions with tails extending in the gap and allowing below-gap absorption at photon energies $h \nu < E_{g,QW}$. (b)~Schematic explaining why type-II transitions have a higher probability in InGaN/GaN structures with low indium content. Band diagrams are calculated with a 1D~Poisson-DD solver and show realistic QW electric fields. (c)~Simulated type-II absorption curves for an In$_{0.11}$Ga$_{0.89}$N/GaN QW as a function of applied bias. (d)~Change in the simulated type-II absorption as a function of bias at a photon energy marked as 'C' in (c).}
\label{fig:schematic_transitions}
\end{figure}

The mechanism at the origin of the the C-region observed in the BPCS measurements at below-gap photon energies is modeled in the following as a type-II well-to-barrier transition, similarly to the process suggested in Ref.~\citenum{Yee2010}. The final state of the transition corresponds to a hole occupying the ground state of the QW and an electron excited in the barrier [Fig.~\ref{fig:schematic_transitions}(a)], which should correspond to the electronic process giving the dominant contribution to the photocurrent signal for type-II absorption because of the light mass of the unbound carrier. Due to the presence of the barrier and QW electric fields both wave functions are Airy functions.

Already at this point we can remark that the present model may explain why the C-region is well observed in UV diodes but progressively disappears going in the visible range. Indeed increasing the indium content of the QW increases the well-barrier conduction band offset $\Delta E_c$ and decreases the threshold photon energy for below-gap transitions. This in turn spatially shifts the final state electron Airy function further from the bound hole wave function, decreasing the e-h overlap and thus the probability amplitude of type-II transitions, as schematized in Fig.~\ref{fig:schematic_transitions}(b). For this reason other mechanisms of below-gap absorption, namely compositional disorder and QCFK in the QW, are observed to dominate in the spectra of the measured solar cells with high indium composition and large well-barrier conduction band offset.

The final electron state $\psi_e$ is calculated in the bulk approximation as an Airy function with parameters defined by the barrier material:\cite{Miller1986}
\begin{align}
\psi_e\left(E_e,z\right) = ~& Ai\Big[\left(\frac{2m_e \cdot e \cdot |\vec{E}_b|}{\hbar^2}\right)^{\frac{1}{3}} \nonumber\\
&\cdot \left(-\frac{E_e}{|\vec{E}_b|} + z \right) \Big] \label{eq:psi_e_Airy}
\end{align}
where $\vec{E}_b$ is the electric field in the barrier of the structure, $m_e$ and $E_e$ are the electron effective mass and energy, respectively, and $z$ is the growth direction. The hole ground state $\psi_{h,0}$ is calculated analytically from the Schr\"odinger equation in the approximation of an infinitely deep potential well of width $L_{QW}$ in presence of an electric field $\vec{E}_{QW}$:
\begin{align}
\psi_{h,0}\left(E_{h,0},z\right) = ~& Ai\left[Z_{h,0}\left(E_{h,0} , z \right)\right] \nonumber\\
& + c_{h,0}~Bi\left[Z_{h,0}\left(E_{h,0} , z \right)\right] \label{eq:psi_h_Airy}
\end{align}
where Ai and Bi are Airy functions; $Z_{h,0}\left(E_{h,0},z\right)$ is defined as $\left(2m_h e |\vec{E}_{QW}|/\hbar^2)\right)^{1/3} \cdot \left(\frac{E_{h,0}}{|\vec{E}_{QW}|} + z \right)$; $m_h$ is the hole effective mass; the coefficient $c_{h,0}$ and the hole energy $E_{h,0}$ are obtained from the boundary conditions $\psi_{h,0} \left( E_{h,0},0 \right) = \psi_{h,0} \left(E_{h,0},L_{QW} \right) = 0$. Then the optical absorption, or equivalently the imaginary part of the optical susceptibility ${\rm Im}(\chi)$, will be proportional to the overlap integral $I(E_e,E_{h,0})$ as:
\begin{align}
{\rm Im}(\chi) = \int & I(E_e,E_{h,0})~\delta(E_e + E_{h,0} + E_{g,QW} \nonumber \\
&+ \Delta E_c - h \nu )~dE_e \label{eq:Im_chi}
\end{align}
where $I\left(E_e,E_{h,0}\right) = |\int_0^{L_{QW}} \psi_e^*(E_e,z) \psi_{h,0}(E_{h,0},z)~dz|^2$ with the integral boundaries fixed by the boundary conditions of $\psi_{h,0}$; $E_{g,QW}$ is the energy gap of the QW; $\Delta E_c$ is assumed to be 80\% of the band offset of the heterojunction.

Inserting the electron and hole wave functions obtained from Eq.~(\ref{eq:psi_e_Airy}) and (\ref{eq:psi_h_Airy}) into Eq.~(\ref{eq:Im_chi}) type-II absorption is calculated as a function of applied bias for a single QW structure representative of the measured solar cell with [In]=11\%. It is assumed $E_{g,QW}$=3.23~eV as the experimental value measured at 5~K, and the bias-dependence of $\vec{E}_{QW}$ and $\vec{E}_b$ is obtained from 1D~Poisson-DD simulations of the device.\cite{Wu2015b}

The results of the model are shown in Fig.~\ref{fig:schematic_transitions}(c). A qualitative agreement is observed between the simulated absorption curves and the experimental features of the C-region shown in Fig.~\ref{fig:responsivity3}(a): increasing the reverse bias from +1~V to -4~V the absorption rapidly increases while the slope of the tail decreases. The change in absorption as a function of bias at a fixed photon energy [marked 'C' in Fig.~\ref{fig:schematic_transitions}(c)] is plotted in Fig.~\ref{fig:schematic_transitions}(d). Qualitatively, the nearly exponential increase in absorption resembles the experimental increase in responsivity measured at 5~K shown in Fig.~\ref{fig:responsivity3}(d). The larger signal observed in the experiments at 300~K in the $\sim$[-1,+1]~V range is attributed to additional mechanisms of below-gap absorption exciting carriers in the QW, namely disorder and FK. Indeed, while type-II transitions directly create carriers in the barrier ready to contribute to the photocurrent signal (if we neglect subsequent recapture by other QWs), other mechanisms involving absorption in the QW will require thermal energy in order to allow thermionic emission from the QW and contribute to photocurrent. These different processes are schematized in Fig.~\ref{fig:responsivity3}(e).

The proposed model gives considerable qualitative insights into the mechanism responsible for the observation of the C-region. Finally, we remark that while the model only considers type-II well-to-barrier transitions, in the experimental C-region it may be present, even at 5~K, a residual contribution to the responsivity signal due to other mechanisms of absorption. This may be the source of quantitative disagreement on the rate of the increase of the modeled absorption vs. the experimental responsivity as a function of bias, in addition to the approximations used ($\psi_e$ in the bulk limit, $\psi_{h,0}$ in the infinitely deep well limit) and the sensitivity of the model on different parameters, such as $\Delta E_c$, $\vec{E}_{QW}$, $\vec{E}_b$, which have been assumed here.


\begin{acknowledgments}
This work has been supported by the project CRIPRONI (ANR-14-CE05-0048-01, MOST-104-2923-E-002-004-MY3, MOST-105-2221-E-002-098-MY3) of the French National Research Agency (ANR) and Taiwanese Ministry of Science and Technology (MOST), and UCSB DARPA Program, DOE SSL. Funding for Bastien Bonef and Robert M. Farrell was provided by the Solid State Lighting \& Energy Electronics Center. Additional support for Marco Piccardo, James S. Speck and Claude Weisbuch was provided by the DOE Solid State Lighting Program under Award \# DE-EE0007096
\end{acknowledgments}

\bibliography{localization}

\end{document}